\documentclass[aps,prd,twocolumn,showpacs,groupedaddress,nofootinbib]{revtex4-1}

\usepackage{graphicx}
\usepackage{dcolumn}
\usepackage{bm}
\usepackage{url}
\def\araa{ARA\&A}%
\def\apj{ApJ}%
\def\apjl{ApJ}%
\def\apss{Ap\&SS}%
\def\aap{A\&A}%
\def\aaps{A\&AS}%
\def\icarus{Icarus}%
\def\mnras{MNRAS}%
\def\prd{Phys.~Rev.~D}%
\def\nat{Nature}%
\begin{document}

\title{Lower-twin-peak quasiperiodic oscillation coherence in x-ray binaries and matter stretched by tides falling onto a compact object}

\author{C. German\`a}
\email{claudio.germana@gmail.com}
\email{claudio.germana@ufma.br}
\affiliation{Departamento de F\'isica, Universidade Federal do Maranh\~ao, 
S\~ao Lu\'is, MA, Brazil}

\date{\today}

\begin{abstract}

Low mass x-ray binaries (LMXBs), with either a neutron star (NS) or a black hole, show in their power
 spectra quasiperiodic oscillations (QPOs). Those at highest frequencies show up in pairs 
 and are named twin peak high frequency QPOs (HF QPOs). Their central frequencies are typical of 
 the orbital motion timescale close to the compact object. HF QPOs are believed to carry unique information on 
 the matter moving in the extreme gravitational field around the compact object. In previous works we highlighted  
 the work done by strong tides on clumps of plasma orbiting in the accretion disk as suitable    
 mechanism to produce the HF QPOs.
 We showed that the upper of the twin peak HF QPOs seen in NS LMXBs could originate from the tidal
 circularization of the clump's relativistic orbit, while the lower HF QPO could come from the spiraling clump
 losing orbital energy. Here we focus on the tidal deformation of a magnetized clump of plasma once tides load 
 energy on it. The likely evolution of the shape of the clump after tidal circularization of the orbit and
 its subsequent orbital evolution are investigated.
In atoll NS LMXBs, a subclass of NS LMXBs 
 less luminous than Z NS LMXBs, the lower HF QPO displays a characteristic behavior 
 of its coherence $Q$ versus its central frequency $\nu$. $Q$ keeps increasing over the range of frequencies $\nu\sim$ 600-850 Hz 
 and then drops abruptly at $\nu\sim$ 900 Hz.
 We note, for the first time, that such behavior is reproduced by magnetized clumps of plasma stretched
 by strong tides and falling onto the NS on unstable orbits. The increasing part of $Q$ is drawn by the tidal stretching 
 timescale of the clump all over the orbit. The abrupt drop of $Q$ is dictated by the number of turns the clump makes 
 before reaching the innermost stable bound orbit; afterwards the clump would fall onto the NS.
 We emphasize the overall behavior of the lower HF QPO coherence as candidate to disclose the innermost stable
 bound orbit predicted by the general relativity theory in the strong field regime.

\end{abstract}


\maketitle

\section{Introduction}\label{sec1}

Low mass x-ray binaries (LMXBs) host a compact object, either
 a neutron star (NS) or a solar-mass black hole (BH), and an evolved star as a companion. 
 The extreme gravitational pull by the compact object captures matter from the evolved star. Matter 
 spirals around the compact object and forms an accretion disk \cite{2002apa..book.....F}.
 The inner part of the accretion disk is expected to lie at orbital radii close to the compact 
 object\footnote{For a maximally spinning BH the innermost stable bound orbit lies at 
 $r=r_{g}$, for a Schwarzschild compact object at $r=6\ r_{g}$. 
 $r_{g}=GM/c^{2}$ is the gravitational radius. $G$ gravitational constant,
 $M$ mass of the compact object and $c$ speed of light. For a 2 M$_{\odot}$ neutron star $r_{g}\sim 3$ km.},
 $r= 1-6\ r_{g}$ \cite{misner}. The plasma in the inner part of the accretion disk orbits in an  
 extreme gravitational field. Thus, LMXBs could be potential laboratories for  
 testing the predictions of general relativity (GR) in the strong field limit, $r\sim r_{g}$ \cite{2003ASPC..308..221L}.

In 1996 the Rossi X-ray Timing Explorer satellite (RTE; \cite{1993A&AS...97..355B}) 
 reported the first submillisecond x-ray brightness oscillations in a NS LMXB \cite{1996ApJ...469L...1V}. 
 Such oscillations were detected in pairs and named twin peak kilohertz quasiperiodic oscillations (kHz QPOs).
 QPOs in LMXBs were long known but at much lower frequencies \cite{1985Natur.317..681L}.
 They range from millihertz to kilohertz and are characterized 
 by their central frequency $\nu$, root-mean-square (rms) amplitude and coherence $Q$ 
 ($Q=\nu/\Delta\nu$ with $\Delta\nu$ full width at
 half maximum of the peak \cite{2004astro.ph.10551V}). BH LMXBs show twin peak QPOs
 at high frequency as well (hundreds of hertz  
 \cite{2006ARA&A..44...49R,2016ASSL..440...61B}), named high frequency (HF) QPOs. In the 
 following we will refer to the pairs observed in both systems as HF QPOs, i.e. either with a NS or a BH, 
 though twin peak kHz QPOs in NS LMXBs show different rms and $Q$  
 than twin peak HF QPOs seen in BH LMXBs \cite{2004astro.ph.10551V}.  
   
The central frequency $\nu$ of HF QPOs is typical of the orbital motion
 timescale (milliseconds) close 
 to the compact object. It is believed that HF QPOs could carry 
 information on the matter orbiting in the inner part of the accretion disk
 \cite{2006AdSpR..38.2675V} and, therefore, be probes to test features of
 the gravitational field around a compact object.
 GR theory states that the space-time around a NS or a BH is 
 strongly curved and this implies (i) the existence of an innermost stable bound
 orbit (ISBO), below which no stable orbital motion takes place
 \cite{misner,1990ApJ...358..538K}, 
 (ii) the periastron precession of the orbits, at a rate of the order of few milliseconds. This 
 would imply modulations of the radiation emitted by orbiting matter
 (in addition to Keplerian modulations) \cite{1999PhRvL..82...17S}, 
 (iii) a relativistic frame dragging by the spinning 
 compact object \cite{1998ApJ...492L..59S}, known as Lense-Thirring precession 
 \cite{1918PhyZ...19..156L}, (iv) moreover, being an extremely curved space-time,
 strong tides by the compact object acting on the orbiting matter
 might play a relevant role. 
   
Despite the dense literature on the subject there is not yet a general agreement on
 the physical mechanism producing the HF QPOs.
 The efforts done to interpret their central frequency have produced several works.
 We mention the main references of each proposed model. 
 The models are based on orbital motion of 
 clumps of plasma in the accretion disk like   
 (i) beat-frequency modulation mechanisms between the clumps
 and the beam of radiation from the spinning NS
 \cite{1998ApJ...508..791M,2001ApJ...554.1210L}, (ii) relativistic orbital motion
 of the clumps modulating the x-ray flux at the Keplerian $\nu_{k}$, the periastron
 precession $\nu_{p}$ and nodal precession $\nu_{nod}$
 frequency of the orbit \cite{1999PhRvL..82...17S,1999ApJ...524L..63S},
 (iii) modulation mechanisms based on resonances between the relativistic frequencies 
 of the clump orbiting in the curved space-time
 \cite{2001A&A...374L..19A,2006AIPC..861..786T}. The models are based on disk oscillations  
 like (iv) diskoseismology \cite{2001ApJ...559L..25W,2012ApJ...752L..18W} and (v) 
 oscillating tori \cite{2016MNRAS.457L..19T,2018MNRAS.474.3967D}.\\ 
In Ref.~\cite{2004ApJ...606.1098S} the signal 
 emitted by a clump of matter orbiting in the accretion disk 
 as seen by a distant observer was modeled in great detail. 
 The authors ray traced in the Kerr metric 
 the photons emitted by an orbiting rigid hot spot. Also, the simulations of the signal coming from an orbiting arc shared
 along the orbit are shown. The simulated power spectra are like those observed in LMXBs. 
 They are characterized by peaks at frequencies corresponding to
 $\nu_{k}$, $\nu_{p}=\nu_{k}-\nu_{r}$, $\nu_{k}+\nu_{r}$ and 
 their harmonics\footnote{The relativistic radial
 frequency $\nu_{r}$ is the number of cycles per second done by a test particle 
 from the periastron of the orbit to apoastron and back to periastron.
 In a curved space-time $\nu_{k}>\nu_{r}$ and this implies the
 periastron precession of the orbit at the frequency $\nu_{p}$.}. 
The detectability that such modulations would have with current and future
 satellites was studied in Ref.~\cite{2014MNRAS.439.1933B}. 
 In both Refs.~\cite{2004ApJ...606.1098S,2014MNRAS.439.1933B} the HF QPOs  
 are produced by relativistic
 effects on the photons emitted by the hot spot,
 such as Doppler boosting and gravitational lensing. These relativistic phenomena
 modulate only a fraction of the radiation emitted by the hot spot.
 For a hot spot of radius $R\sim 0.5\ r_{g}$, as bright as twice
 the background radiation from the disk, the modulations are of the 
 order of some percents \cite{2004ApJ...606.1098S}.\\
Ray tracing in the Schwarzschild metric of the photons emitted by a not-rigid
 sphere is presented in Refs.~\cite{2005PhRvD..72j4024C,2009A&A...496..307K,2010AIPC.1205...30C}.
 Such simulations reproduce the effects of tidal deformation of the sphere on the 
 signal collected by the distant observer. As long as the sphere orbits it is squeezed
 and elongated by tides into an arc along the orbit. The numerical code was proposed to fit
 the near infrared-x-ray flares observed 
 at the galactic center \cite{2003Natur.425..934G}. In Ref.~\cite{2009A&A...496..307K} the flares
 are produced by small satellites (like asteroids or comets) captured by the strong gravitational 
 field of the supermassive BH at the Galactic center. The strong tidal
 force by the BH disrupts the small satellite and thus some percent of its
 rest-mass energy is emitted as radiation. The numerical code was used to calculate
 the power spectrum of the signal. Power spectra reproducing  
 those observed in LMXBs were obtained: the power law with 
 superimposed the twin peak 
 HF QPOs at the frequencies $\nu_{k}$ and $\nu_{k}+\nu_{r}$ are reproduced 
 \cite{2009AIPC.1126..367G}. The power law is produced
 because of the abrupt increase of the luminosity of the clump during tidal
 stretching \cite{2009A&A...496..307K}. The twin peaks are produced by Doppler 
 boosting of light and gravitational lensing.\\
 Ref.~\cite{2013MNRAS.430L...1G} highlighted the likely root mechanism 
 that produces multiple peaks in the power spectrum for matter orbiting
 in a curved space-time. The azimuthal phase 
 $\phi(t)$ of a body orbiting on a slightly eccentric orbit in a curved
 space-time is not a simple linear function of time with slope $\nu_{k}$. 
 The phase also oscillates at the relativistic radial frequency $\nu_{r}$.
 In a flat space-time $\nu_{k}=\nu_{r}$ 
 while in a curved space-time $\nu_{r}<\nu_{k}$. This implies that in the power 
 spectrum of $\phi(t)$ the peaks $\nu_{k}$ and $\nu_{k}\pm\nu_{r}$ are produced. 
 The timing law $\phi(t)$ is transformed into an observable
 light curve by Doppler boosting of light. Furthermore, the fact that the clump of matter
 is stretched by tides may cause some peaks not be seen in the power spectrum 
 \cite{2009AIPC.1126..367G}. 
 Only one source has showed evidence for a triplet of peaks, 
 BH LMXB XTE J1550-564 \cite{2002ApJ...580.1030R}. 
 However, the peak at lowest frequency 
 (corresponding to $\nu_{p}=\nu_{k}-\nu_{r}$ in this framework and not
 seen in the simulation \cite{2009AIPC.1126..367G}) was marginally detected.
 The best case for multiple peaks detection is the BH LMXB GRS 1915+105 
  \cite{2001A&A...372..551B,2001ApJ...554L.169S,2006ARA&A..44...49R,2013MNRAS.432...19B}.
 
The work done by the strong tidal force
 on clumps of matter orbiting close to a compact object could be a 
 source of energy that might justify how the HF QPOs would originate.
 The orbiting body needs to be overbright 
 with respect to the accretion disk in order to produce detectable modulations 
 \cite{2004ApJ...606.1098S}. Thus, the release of gravitational energy by the
 clump because of tides might be a valid ingredient to interpret   
 where the energy carried by the HF QPOs (related to their rms amplitude)
 comes from. Tidal disruption events (TDEs) in which stars are disrupted 
 by supermassive BHs at the center of galaxies,
 releasing large amounts of energy as a flare, have already been discovered 
 (e.g. Refs.~\cite{2014ApJ...783...23G,2015Natur.526..542M,2015JHEAp...7..148K,2017MNRAS.468..783L}).
 It is worth mentioning that Ref.~\cite{2012Sci...337..949R} reports a QPO
 detected in the x ray flux coming from the tidal disruption of a star.
 In our Solar System we have proofs that the tidal force can extract
 significant amounts of energy. The intense volcanism of Jupiter's
 moon Io \cite{1979Sci...203..892P}, likely the plumes from Jupiter's moon 
 Europa \cite{2018NatAs...2..459J} and Saturn's moon Enceladus  
 \cite{2008Icar..194..675R,2014Sci...344...78I,2006Sci...311.1422H,2014Icar..235...75S}.

The characteristic behavior of both the 
 rms amplitude and coherence $Q$ of HF QPOs was reported in
 Refs.~\cite{2001ApJ...561.1016M,2006MNRAS.370.1140B,2006MNRAS.371.1925M} 
 for several NS LMXBs. In atoll NS LMXBs \cite{1989A&A...225...79H} the amplitude 
 of the lower HF QPO increases and then decreases as a function of the central 
 frequency $\nu$ of the peak. Its coherence $Q$ increases and then drops abruptly. 
 Such behavior was highlighted \cite{2006MNRAS.370.1140B}
 as a possible candidate to disclose the ISBO predicted
 by GR, below which no stable orbital motion exists. 
 The rms of the upper HF QPO keeps decreasing as a function of the central frequency 
 $\nu$ of the peak. Its coherence displays an almost flat trend of the order 
 of $Q\sim 10$ over a large range frequencies.\\      
If the HF QPOs originate from clumps of plasma orbiting
 in the accretion disk, by attempting to model the rms and $Q$,
 we could extract some information about 
 the gravitational field around the compact object. 
 A modeling of such trends is presented in Refs.~\cite{2015PhRvD..91h3013G,2017PhRvD..96j3015G} 
 (hereafter GC15 and G17, respectively). Making use of the gravitational potential
 in the Schwarzschild metric we showed that the energy extracted by tides 
 from clumps of plasma orbiting in the accretion disk 
 might produce the rms amplitude of the HF QPOs seen in NS LMXBs. 
In GC15 we showed that the amplitude of the lower HF QPO seen in NS LMXBs 
 could come from the energy released by the spiraling clump because of the removal of 
 orbital energy by tides. In G17 we focused on the upper HF QPO and showed that both 
 its amplitude and coherence are in agreement 
 with those produced by tidal circularization of the clump's relativistic orbit.
 We also derived a magnetic field of the clump typical of 
 that in atoll NS LMXBs ($B\sim10^{8}-10^{9}$ G 
 \cite{1999A&AT...18..447P}) and highlighted that the deformation of the clump by 
 strong tides, and consequent perturbation of $B$, might set the mechanism 
 to turn the released orbital energy into radiation seen as HF QPOs
 \cite{2009A&A...496..307K}. Such mechanism could be 
 synchrotron emission by the plasma electrons spiraling around the magnetic field lines. The 
 orbital energy extracted by tides from the clump is loaded as internal energy 
 \cite{1977ApJ...213..183P,2009A&A...496..307K} of the clump. As long as 
 the clump is squeezed by tides its magnetic field increases and the energy
 might go into kinetic energy of the electrons. 
 It is interesting to mention that in Ref.~\cite{2013MNRAS.434..574S} has been shown 
 that the hard x-ray radiation observed in two x-ray binaries, 
 over 10-100 milliseconds time intervals, could originate
 from cyclosynchroton self-Compton mechanisms. Simulations in 
 Ref.~\cite{2017MNRAS.469.4879B} show that a star magnetic field  
 stretched by strong tides increases at least by a factor of 10.
 
In this manuscript we aim to show how a clump of plasma, of magnetic field $B$, would react  
 to tidal pressure during tidal circularization of its relativistic orbit.
 We investigate the evolution of the shape of the clump and its subsequent phase,
 i.e. the clump deformed by tides into a prolate spheroidal-like object  
 orbiting on a circular orbit and undergoing further tides. Our purpose is to estimate 
 the temporal coherence of the Keplerian modulation produced to compare it to the 
  twin peak HF QPOs coherence seen in atoll NS LMXBs 
 \cite{2006MNRAS.370.1140B,2006MNRAS.371.1925M}.\\   
The manuscript is organized as follows. In Sec.~\ref{sec2} we recall the main arguments 
 and results presented in both GC15 and G17. In Sec.~\ref{sec3} we show how 
 a magnetized clump of plasma would deform under tides.  
 The initial spherical clump orbits on a slightly eccentric orbit in the Schwarzschild metric, 
 later circularized by tides. 
 We estimate the cross-section radius and polar axis of the 
 clump deformed into a prolate spheroid, once tidal deformation
 stops because internal magnetic pressure.  
 In Sec.~\ref{sec4} we study the subsequent evolution phase, i.e. 
 the spheroid orbiting on a circular orbit and undergoing more tides. 
 In Sec.~\ref{sec5} we summarize the conclusions.   
       
\section{Tidal load and orbital energy of clumps of plasma}\label{sec2}

It is worth mentioning that magnetohydrodynamics simulations show 
 an inner part of the accretion disk highly turbulent
 \cite{2001ApJ...548..348H}. Furthermore, the discovery of large structures
 in the accretion disk of a x-ray binary was reported in Ref.~\cite{2013Sci...339.1048C}.
 Propagating accretion rate fluctuations in the disk are modeled
 \cite{2013MNRAS.434.1476I,2016AN....337..385I} 
 to reproduce the aperiodic variability observed in BH LMXBs.  
 Therefore, it is reasonable to think an accretion disk is characterized
 by inhomogeneities propagating throughout it.
 Ref.~\cite{1989Ap&SS.158..205H} noted that magnetically confined
 massive clumps of plasma might form in the inner part of the accretion disk. 

In Ref.~\cite{2009AIPC.1126..367G} the power spectrum of the signal from  
 tidal disruption of a clump of matter by a Schwarzschild black hole is shown. The 
 simulated power spectrum reproduces the power spectra observed in LMXBs. 
 The simulation shows a power law with superimposed twin peak HF QPOs.
 Motivated by this result, in GC15 we calculated the energy that would be
 released by a clump of plasma spiraling in the accretion disk in LMXBs,
 when the clump orbital energy is removed by tides. We highlighted there that the 
 lower HF QPO in atoll NS LMXBs could be the Keplerian modulation produced by orbiting matter. 
 In G17 we instead focused on the physical mechanism that would produce the upper HF QPO. 
 We showed that the energy released by the clump, during tidal circularization of 
 its slightly eccentric relativistic orbit, accounts for both the amplitude and coherence of the 
 upper HF QPO seen in atoll NS LMXBs \cite{2006MNRAS.370.1140B,2006MNRAS.371.1925M}.
 The tidal evolution of the orbits of a low-mass satellite orbiting a Schwarzschild BH have been 
 studied in detail in Ref.~\cite{2008A&A...487..527C}, showing that inner orbits circularize and 
 shrink. Below we briefly recall the main arguments and results presented in both GC15 and G17.
 
A spherical clump of radius $R$ undergoes a tidal force  
 (see also GC15)
\begin{eqnarray}\label{eq1}
F_{T}&=&\mu'c^{2}\left[\left(\frac{dV_{eff}}{dr}\right)_{(r-R)}-\left(\frac{dV_{eff}}{dr}\right)_{(r+R)}\right]\\ \nonumber
&\approx&\mu'c^{2}2R\left(\frac{d^{2}V_{eff}}{dr^{2}}\right)_{r}
\end{eqnarray} 
 where $\mu'=\rho V'$ is the mass of the spherical cap of the clump, of height, say, one-tenth
 of the radius $R$ of the clump, $h=R/10$ 
 (as in GC15 and G17), $\rho$ the density of the material. 
 The cap has a volume $V'= \pi h^{2}(R-h/3)$. In (\ref{eq1}) $V_{eff}$ is the 
 effective gravitational potential in the Schwarzschild metric 
\begin{equation}\label{eq2}
 V_{eff}=1-\frac{2m}{r}-\frac{2m\tilde{L}^{2}}{r^{3}}+\frac{\tilde{L}^{2}}{r^{2}}
 \end{equation} 
with $m$ the mass of the compact object in geometric units and $\tilde{L}$ the angular momentum per unit mass
 of a test particle, orbiting on an orbit of semilatus rectum $p$ and eccentricity $e$,
\begin{equation}\label{eq3}
\tilde{L}\left(m,p,e\right)=\left(\frac{p^{2}m^{2}}{p-3-e^{2}}\right)^{1/2}
\end{equation}
 $p$ is linked to the periastron of the orbit through $r_{p}=pm/(1+e)$ \cite{1994PhRvD..50.3816C}.

For a solid-state clump of matter, electrochemical bounds keep the clump together. This 
 internal force is characterized by the ultimate tensile strength $\sigma$ of the material, i.e. 
 the internal force per unit area. The clump of matter, in order not to be broken into smaller pieces by tides,  
 should have internal forces larger than the tidal force, $2\pi R h\sigma\ge F_{T}$.
 We can get some order of magnitude
 on the maximum radius $R_{max}$ of the clump set by tides
\begin{eqnarray}\label{eq4}
R_{max}&=&\left(10\left(1-\frac{1}{30}\right)^{-1}\frac{c_{s}^{2}}{c^{2}}\frac{\sigma}{Y}\times\right.\\ \nonumber
&&\left.\left(-\frac{2m}{r^{3}}+\frac{3\tilde{L}^{2}}{r^{4}}-\frac{12m\tilde{L}^{2}}{r^{5}}\right)^{-1}\right)^{1/2}
\end{eqnarray}
 where we wrote the density $\rho=c_{s}^2/Y$; $Y$ is Young's modulus of the material, $c_{s}$ the 
 speed of sound in it. In  
Sec. IV of GC15 we treated a clump of plasma   
 in the accretion disk around LMXBs as characterized by some internal force per unit area $\sigma$ 
 (e.g. electrochemical bounds and/or a magnetic force). The speed of sound in (\ref{eq4})  
 for the plasma case comes from the standard accretion disk model \cite{2002apa..book.....F} 
\begin{equation}\label{eq5}
c_{s}=\left(\frac{\gamma Z k T(r)}{m_{i}}\right)^{1/2}    
\end{equation}
with $T(r)$ the temperature profile from the equations of the standard disk \cite{2002apa..book.....F},
 $\gamma\sim5/3$ is the adiabatic index, $Z$ the charge state ($Z=1$ for
 a hot plasma), $m_{i}$ the ion hydrogen mass, $k$ Boltzmann's constant \cite{2002apa..book.....F}.\\ 
Using the energy observed in HF QPOs ($\sim 10^{36}$ erg/s), in GC15 we derived the ratio 
 $\sigma/Y$ in (\ref{eq4}), $\sigma/Y\sim 300$ and $\sim 70$
 for atoll and Z NS LMXBs, respectively. In G17 we argued that the pressure 
 $\sigma$ keeping the clump together is consistent with the magnetic pressure because
 of the NS magnetic field, which might permeate the clump in the disk.
 We obtained a magnetic field permeating the clump consistent with that estimated in NS LMXBs
 ($B\sim 10^{8}-10^{9}$ G \cite{1999A&AT...18..447P}). Note that in Ref.~\cite{1989Ap&SS.158..205H}
 it is emphasized that magnetically confined massive clumps of plasma can form in the
 inner part of the accretion disk. Therefore, following these results, in this manuscript 
 we aim to study the deformation of the magnetized clump of plasma, when tides load 
 energy against the internal magnetic pressure of the clump.
 To take into account that the magnetic field permeating the plasma in the
 accretion disk might decrease with distance from the NS, 
 here we set a magnetic momentum of the NS $\mu=5\times 10^{27}$ G/cm$^{3}$.
 Thus, for a dipolar magnetic field, the magnetic field decreases as
 $B=\mu/r^{3}\sim 8\times 10^{8}-2\times 10^{8}$ G
 over the range of radii $r\sim 6-10\ r_{g}$, such that we are consistent  
 with the $B$ estimated in G17 from the equation for $\sigma/Y$ and using the energy observed in HF QPOs.
 The internal magnetic pressure (in Pascal) of the clump of plasma is $p_{m}=B^{2}/2\mu_{0}=\sigma$, with $\mu_{0}=4 \pi \times 10^{-7}$
 H/m vacuum magnetic permeability.
 
In CG15 and G17 we mentioned that clumps with $R=R_{max}$ from (\ref{eq4}) do not probably form at all 
 because of the strong tidal force. The tidal load on the clump (the tidal force (\ref{eq1}) per unit area)
 has to be $n$ times smaller than $\sigma$, i.e. $F_{T}/2\pi R h=\sigma_{T}=\sigma/n$, where
\begin{eqnarray}\label{eq6}
\sigma_{T}&=&\frac{\mu'c^{2}}{2\pi Rh}\left[\left(\frac{dV_{eff}}{dr}\right)_{(r-R)}-\left(\frac{dV_{eff}}{dr}\right)_{(r+R)}\right]\\ \nonumber
&\approx&\frac{10\mu'c^{2}}{\pi R}\left(\frac{d^{2}V_{eff}}{dr^{2}}\right)_{r}
\end{eqnarray}
In GC15 and G17 we derived $n=5$ as upper limit, which gives $R\sim3000$ m over a range of orbital radii 
 $r\sim 6-10\ r_{g}$ (see Fig.~1 in G17). 
In G17 (Fig.~3) we estimated the energy emitted by the clump of plasma in order to circularize its 
 slightly eccentric orbits, of initial eccentricity\footnote{We already mentioned in G17
 that because of the turbulent environment in an accretion disk \cite{2001ApJ...548..348H},
 it may be reasonable thinking that clumps may orbit on not perfect circular orbits. Power spectra much like 
 those observed in LMXBs are obtained for orbit with low eccentricity $e=0.1$ \cite{2004ApJ...606.1098S,2009AIPC.1126..367G,2013MNRAS.430L...1G,2014MNRAS.439.1933B}.}
 $e=0.1$. This energy is given by 
\begin{equation}\label{eq7}
\epsilon=\mu c^{2}\left(\tilde{E}\left(p,e\right)-\tilde{E}\left(p,0\right)\right)
\end{equation} 
 where $\mu$ is the total mass of the clump and $\tilde{E}(p,e)$ is the orbital energy 
 (in geometric units) per unit mass in the Schwarzschild metric \cite{1994PhRvD..50.3816C}
\begin{equation}\label{eq8}
\tilde{E}\left(p,e\right)=\left(\frac{\left(p-2-2e\right)\left(p-2+2e\right)}{p\left(p-3-e^{2}\right)}\right)^{1/2}
\end{equation}

\section{Tidal deformation of magnetized clumps of plasma}\label{sec3}
   
In G17 we have seen that the work done by tides on clumps orbiting at different orbital radii,
 removing the orbital energy at each periastron passage, gives timescales of tidal circularization of the
 orbits in agreement with the coherence factor $Q$ of the upper HF QPO seen in atoll NS LMXBs \cite{2006MNRAS.370.1140B}.
 Moreover, the energy released by the clump over such timescale, and modulated by Doppler boosting 
 \cite{2004ApJ...606.1098S}, agrees to the upper HF QPO amplitude from the observations \cite{2006MNRAS.370.1140B}. Therefore, 
 we concluded that the upper HF QPO might unveil the tidal circularization of relativistic orbits occurring 
 around a NS. 

Like in G17 the tidal load (\ref{eq6}) is integrated over
 one periastron passage to get the load per periastron passage. We estimate the timescale the tidal
 wave propagates all over the clump of radius $R\sim 3000$ m 
 as $t=2R/c_{s}\sim 0.01$ s, where $c_{s}\sim 5\times 10^{5}$ m/s is the speed of sound in the plasma from (\ref{eq5}). 
 Because the timescale of tidal circularization from G17 is of the same order ($\sim 0.01$ s), over the range of 
 orbital radii $r\sim 6-10\ r_{g}$, in first approximation we can treat the clump as having a spherical
 shape during all the periastron passages needed to circularize the orbit ($N\sim$2-5). Therefore, the total tidal pressure 
 $p_{T}$ deposited on the clump is the tidal pressure per periastron passage
 (i.e. $\sigma_{T}$ integrated over one periastron passage) times the 
 number of periastron passages. 
\begin{figure}[!t!]
\includegraphics[width=0.45\textwidth]{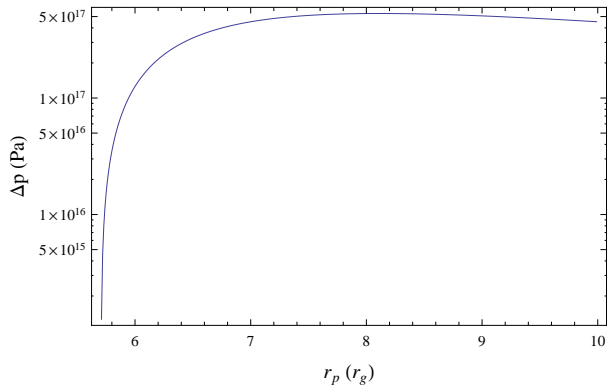}
\caption{The difference between tidal pressure $p_{T}$ loaded on a clump of 
 plasma during tidal circularization of its orbit 
 and internal magnetic pressure of the clump $p_{m}$, 
 as a function of the periastron of the orbit around a $2\ M_{\odot}$ Schwarzschild compact object.}\label{fig1}
\end{figure}
Fig.~\ref{fig1} shows the difference $\Delta p$ between the 
 total tidal pressure $p_{T}$ deposited on the clump during tidal circularization
 and the internal magnetic pressure $p_{m}$ of the clump. The difference 
 drops close to ISBO ($r_{p}\sim 5.6\ r_{g}$ for $e=0.1$) because  
 the energy released to circularize the orbit decreases approaching ISBO (G17).
 Therefore, the tidal energy loaded on the clump during circularization is lower. This is 
 a consequence of the flattening of the gravitational potential minimum close to ISBO, because 
 of the term $\propto 1/r^{3}$ in (\ref{eq2}) \cite{misner}.

Numerical simulations show that tidal deformation squeezes and elongates 
 the clump into a polelike object along the orbit \cite{2010AIPC.1205...30C}.
 The clump is squeezed in the directions perpendicular to the orbital motion
 and elongated along the orbital motion. Since the tidal tensor is traceless it implies 
 volume conservation.\\ 
 The electric conductivity is extremely high in plasmas and 
 we could say, in first approximation, that the magnetic field $B$ of the clump is frozen.
 Following Alve\'en's theorem this leads to magnetic flux conservation. Therefore, the magnetic field
 probably keeps increasing as long as the clump is squeezed by tides \cite{2009A&A...496..307K}. 
 Recent numerical simulations of a star magnetic 
 field squeezed by tides indeed demonstrated that the magnetic field increases at least by a factor of 
 $\sim$ 10 \cite{2017MNRAS.469.4879B}. Squeezing of the clump of plasma by tides goes on probably until the  
 internal magnetic pressure of the increasing magnetic field equals the total tidal pressure 
 deposited on the clump. Eventually, the initial spherical clump is deformed
 by tides into a prolate spheroidal-like object of cross-section radius $R'$ and semimajor axis $a$. 
 The shape of an object deformed by 
 tides, orbiting a Schwarzschild BH, was studied by means of numerical simulations in 
 Refs.~\cite{2009A&A...496..307K,2010AIPC.1205...30C}, 
 however, without taking into account the internal pressure of the clump. Here we consider both internal
 magnetic pressure $p_{m}$ and external tidal pressure $p_{T}$. Because of the complexity of the problem, 
 we attempt to give a general picture on how the clump would deform under tides.

The increase $\Delta B$ of the magnetic field because of tidal compression is equal to 
\begin{equation}\label{eq9}
\Delta B=\left(2\mu_{0} \Delta p\right)^{1/2}
\end{equation}
where $\Delta p$ is shown in Fig.~\ref{fig1}. We find that $\Delta B\sim 5\times 10^{9}$ G and obviously reflects 
 the behavior of Fig.~\ref{fig1}, dropping to $\Delta B\sim 10^{7}$ close to ISBO. Thus, we might say that 
 during tidal circularization of the orbit the magnetic field of the clump has increased by a factor of 
 10, i.e. $B+\Delta B=B'\sim 5\times 10^{9}$ G over the range $r=6.5-10\ r_{g}$. At ISBO $B'\sim 10^{9}$
 since less energy was loaded on the clump to circularize the orbit. The clump is less squeezed. It is worth remarking that 
 this calculation agrees to the magnetic field of the clump derived in G17, however calculated in a  
 different way, using the energy observed in HF QPOs and equation (11) there.

\begin{figure}[!t!]
\includegraphics[width=0.45\textwidth]{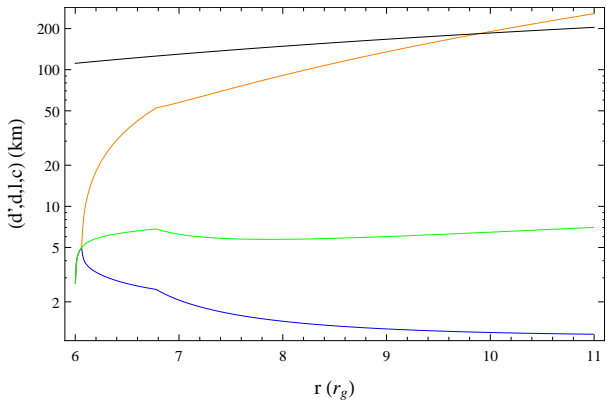}
\caption{Dimensions of the initial spherical and, later deformed by tides, spheroidal clump 
 as a function of the radius $r$ of the circular orbit. At $r\sim 6\ r_{g}$ the 
 orange and blue lines are superimposed
 to the green one. \emph{Blue line}: cross-section diameter $d'$ of the prolate
 spheroid deformed by tides once tidal deformation stops because of
 internal magnetic pressure; \emph{Green line}: diameter $d$ set by tides of the spherical clump before
 tidal deformation (G17); \emph{Orange line}: length of the prolate spheroid once tidal deformation stops;
 \emph{Black line}: circumference of the orbit.}\label{fig2}
\end{figure}

To estimate the dimensions of the clump, turned by tides into prolate spheroidlike object, in first approximation
 we consider the increase of the frozen magnetic field as \cite{2010AIPC.1205...30C} 
\begin{equation}\label{eq10}   
B'=\frac{B S_{in}}{S'}
\end{equation}
with $B'=B+\Delta B$, $S_{in}$ is the initial cross section of the spherical clump,
 $S'$ is the cross section of the prolate spheroid, 
 when tidal pressure equals internal magnetic pressure. From (\ref{eq10}) and 
 volume conservation condition we can derive the cross-section 
 radius $R'$ of the prolate spheroid and its semimajor axis $a$
\begin{equation}
R'=\left(\frac{B R^{2}}{B'}\right)^{1/2}
\end{equation}
\begin{equation}
a=\frac{3V}{4\pi R'^{2}}
\end{equation}
where $V$ is the volume of the initial sphere equal to that of the spheroid $V=4/3 \pi a R'^{2}$.  
Figure \ref{fig2} shows the dimensions of the initial spherical clump and those of the tidally deformed 
 prolate spheroid, after tidal circularization of the orbit.
 We see that close to the innermost stable orbit, now circular at $r=6\ r_{g}$ 
 and known as the innermost stable
 circular orbit (ISCO), the clump is little deformed, because less energy is loaded on it (G17). 
 The black line in the figure
 indicates the circumference of the orbit. The green line 
 is the initial diameter $d=2R$ of the spherical clump set by tides as derived in G17. The blue line is the diameter 
 cross section of the deformed spheroid $d'=2R'$, the orange line its length $l=2a$. 
 We see that the clump is squeezed by tides from $\sim$6 km to $\sim$1-2 km and its length 
 along the orbit, as numerical simulations show \cite{2010AIPC.1205...30C}, goes up 
 to $\sim$ 10-200 km. We note that for orbital radii $r\ge 10\ r_{g}$ the spheroid is elongated by tides 
 all over the circumference of the orbit. 

The results in both Figs.~\ref{fig1},\ref{fig2}
 are for a clump of plasma orbiting around a $2\ M_{\odot}$ neutron star, with orbits of
 initial eccentricity $e=0.1$, later circularized by tides (G17). The luminosity of the
 accretion disk is typical for an atoll NS LMXB
 ($L\sim 0.1 L_{Edd}$, $L_{Edd}\sim 2.5\times 10^{38}$ erg/s
 Eddigton luminosity), giving a density $\rho\sim 1.5$ g/cm$^{3}$ of the plasma, 
 from the equations of the standard accretion disk \cite{2002apa..book.....F}. 

\section{Evolution after tidal circularization of the orbit}\label{sec4}

The squeezed clump of plasma finds itself on a circular orbit. 
 In first approximation, we recalculate the tidal load (\ref{eq6}) with $R'=d'/2$ as in Fig.~{\ref{fig2}} and 
 consider that the clump now has a prolate spheroidal shape along the orbit \cite{2010AIPC.1205...30C}, 
 no longer a spherical one. While in the spherical case we approximated the tidal force as the
 difference of the gravitational force on two radially opposite spherical caps, in the spheroidal case 
 it is the difference on two opposite radial slices of the spheroid: the one at $r-R'$ and that at $r+R'$, 
 of approximated area and volume $A\sim d'l$, $V'\sim d'l h$, respectively, with $h=R'/10$ the thickness of the radial slice.  
 The tidal load (\ref{eq6}) now reads
\begin{eqnarray}\label{eq11}
\sigma_{T}'&=&\frac{\mu'c^{2}}{d'l}\left[\left(\frac{dV_{eff}}{dr}\right)_{\left(r-R\right)}-\left(\frac{dV_{eff}}{dr}\right)_{\left(r+R\right)}\right]\\ \nonumber
&\approx&\frac{\mu'c^{2}}{l}\left(\frac{d^{2}V_{eff}}{dr^{2}}\right)_{r}
\end{eqnarray}
 with $\mu'\sim \rho V'$ the mass of the radial slice. The total tidal load after $N$ Keplerian turns
 on a circular orbit would be $N 2\pi \sigma_{T}'$. At each Keplerian turn tides squeeze and elongate 
 further the spheroid along the orbit, as shown by numerical simulations \cite{2009A&A...496..307K,2009AIPC.1126..367G,2010AIPC.1205...30C}.
 During squeezing radiation could be emitted through synchrotron mechanisms because tidal energy is 
 probably transferred to the electrons in the plasma that spiral around the magnetic field
 lines \cite{2009A&A...496..307K}. Ref.~\cite{2013MNRAS.434..574S} shows 
 that the hard x-ray radiation observed in two x-ray binaries, over 10-100 milliseconds time intervals,
 could come from cyclosynchroton self-Compton emission.

The Keplerian modulation produced by the orbiting spheroid would last until, at least, the spheroid is elongated all over the orbit. 
 Afterwards, the azimuthal asymmetry required to produce a Keplerian modulation no more exists. 
 Equating the length $l'$ of the deformed spheroid after $N$ Keplerian turns to the circumference 
 of the orbit $c=2\pi r$, we can estimate the number of Keplerian turns $N$ needed in 
 order to elongate the spheroid all over the orbit. It reads 
\begin{equation}\label{eq12}
N=\frac{1}{p_{T}}\left[p_{m}+\frac{240 p_{m}}{l^{2}}\left(\pi r-\frac{l}{2}\right)^{2}\right]-\frac{p_{m}}{p_{T}}
\end{equation}
with $p_{m}=B'^{2}/2\mu_{0}$ the internal magnetic pressure of the spheroid after tidal circularization of the orbit,
 $p_{T}=2\pi \sigma_{T}'$ the tidal pressure loaded on it over the first Keplerian turn on the circular orbit.
 The term $p_{m}/p_{T}$ in (\ref{eq12}) takes into    
 account that the internal magnetic pressure of the spheroid,
 because of squeezing during tidal circularization, has increased and
 now, for several initial Keplerian turns ($\sim 5-50$ over the range $r=6-10\ r_{g}$), is higher than
 the external tidal pressure $p_{T}$.
\begin{figure}[!t!]
\includegraphics[width=0.45\textwidth]{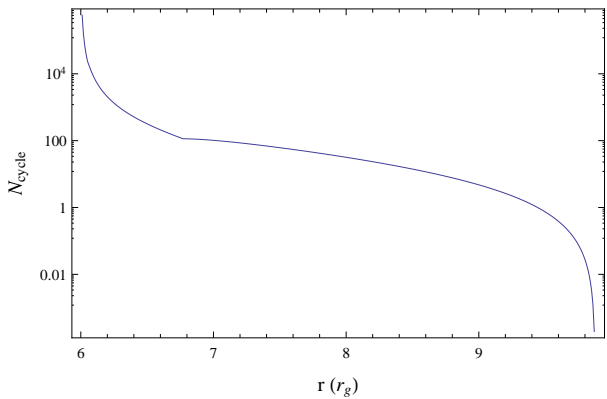}
\caption{Effective number of Keplerian turns to get the spheroid stretched by tides all over the orbit, 
 as a function of the orbital radius.}\label{fig3}
\end{figure}
The tidal force in the case of the spheroid applies on a larger area than the sphere case, thus 
 the tidal pressure weakens. The spheroid on the circular orbit will start be squeezed by tides
 after $N'=p_{m}/p_{T}\sim 5-50$ Keplerian  
 turns. Since within this context radiation is probably emitted by the clump only when it is
 deformed by tides, the temporal coherence 
 of the signal emitted is characterized only by the number of Keplerian turns during tidal squeezing.
 Therefore we take the residuals $N$, subtracting to the total number of turns $N_{tot}$ the turns   
 before squeezing $N'=p_{m}/p_{T}$, when tidal energy
 is loaded and the clump is not deformed since internal magnetic pressure is higher.
 An issue that should be considered is the not instantaneously reaction of the clump
 to external tidal deformation. Most probably the tidal squeezing wave propagates through
 the clump in a finite time $\tau=2 R''/c_{s}$ ($c_{s}$ speed of sound), thus delaying by $\tau$ the deformation 
 induced by tides at each Keplerian turn. Therefore, in first approximation, we might say that the effective
 number of Keplerian cycles 
 the clump to get elongated all over the orbit is $N_{cycle}\sim N (\tau/t_{k})$, with $t_{k}$
 the Keplerian period on the circular orbit.        

Fig.~\ref{fig3} shows the number of Keplerian cycles to get the spheroid elongated
 all over the orbit. As we expect, it is zero for $r\ge 10\ r_{g}$
 since at this radii, as seen in Fig.~\ref{fig2}, the spheroid is elongated all over 
 the orbit already during tidal circularization of the orbit. Therefore, 
 for $r\ge 10\ r_{g}$ Keplerian modulations are produced just during tidal circularization
 of the orbit, with a low coherence $Q\sim 5$ (G17). Instead, we see 
 that for smaller radii the number of turns increases, with the highest
 value at ISCO. The increase of the number of  
 Keplerian cycles towards ISCO is a direct consequence of the weakened tidal force
 towards ISCO. In Fig.~2 of G17 we have already pointed out this feature of a curved space-time  
 and in G15 we have seen that it is caused by the flattening of the gravitational potential minimum, 
 because of the term $\propto 1/r^{3}$ in (\ref{eq2}) \cite{misner}.

There is one more issue that should be taken into account in order to make 
 reliable the results in Fig.~\ref{fig3}. It is known that at each Keplerian turn 
 tides remove orbital energy from an orbiting body, depositing it on the body as internal energy 
 \cite{1977ApJ...213..183P,2008A&A...487..527C}. Therefore, a clump orbiting at ISCO, for example, cannot 
 make many Keplerian turns because it will soon enter into an inner unstable orbit and fall onto the compact object. 
 Perhaps, because of the definition of ISCO, we could expect that a body orbiting at ISCO would produce
 a signal with a number of cycles equal to 0, since it soon falls onto the compact object.
 To take into account this issue we calculate the interval of radii $\Delta r$ over which the spheroid 
 spirals because of the removal of orbital energy by tides, at each Keplerian turn. It comes from equation (\ref{eq8})
\begin{equation}\label{eq15}
\frac{\Delta r}{r_{g}}=\left(\frac{d\tilde{E}}{dp}\right)^{-1}\frac{E_{T}}{\mu c^{2}}\left(\frac{\tau}{t_{k}}\right)^{-1}
\end{equation}  
\begin{figure}[!t!]
 \includegraphics[width=0.45\textwidth]{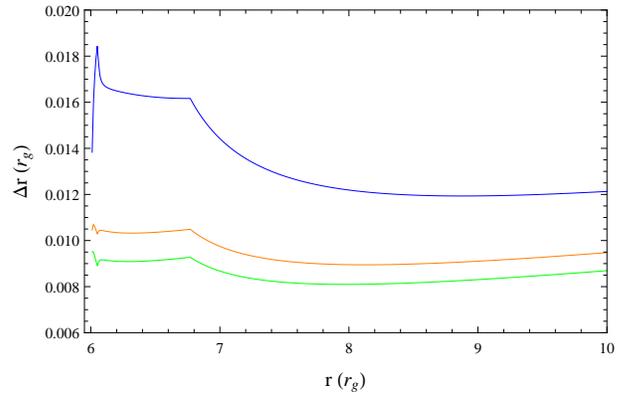}
 \caption{Radial drift $\Delta r$ (in units of $r_{g}\sim$ 3 km) 
 because of the removal of orbital energy by tides within one Keplerian turn
 as a function of the orbital radius $r$.
 \emph{Blue line}: $\Delta r$ after turn 1. \emph{Orange line}: $\Delta r$ 
 over turn 50. \emph{Green line}: $\Delta r$ over turn 100.}.\label{fig4}
\end{figure}
\begin{figure}
\includegraphics[width=0.45\textwidth]{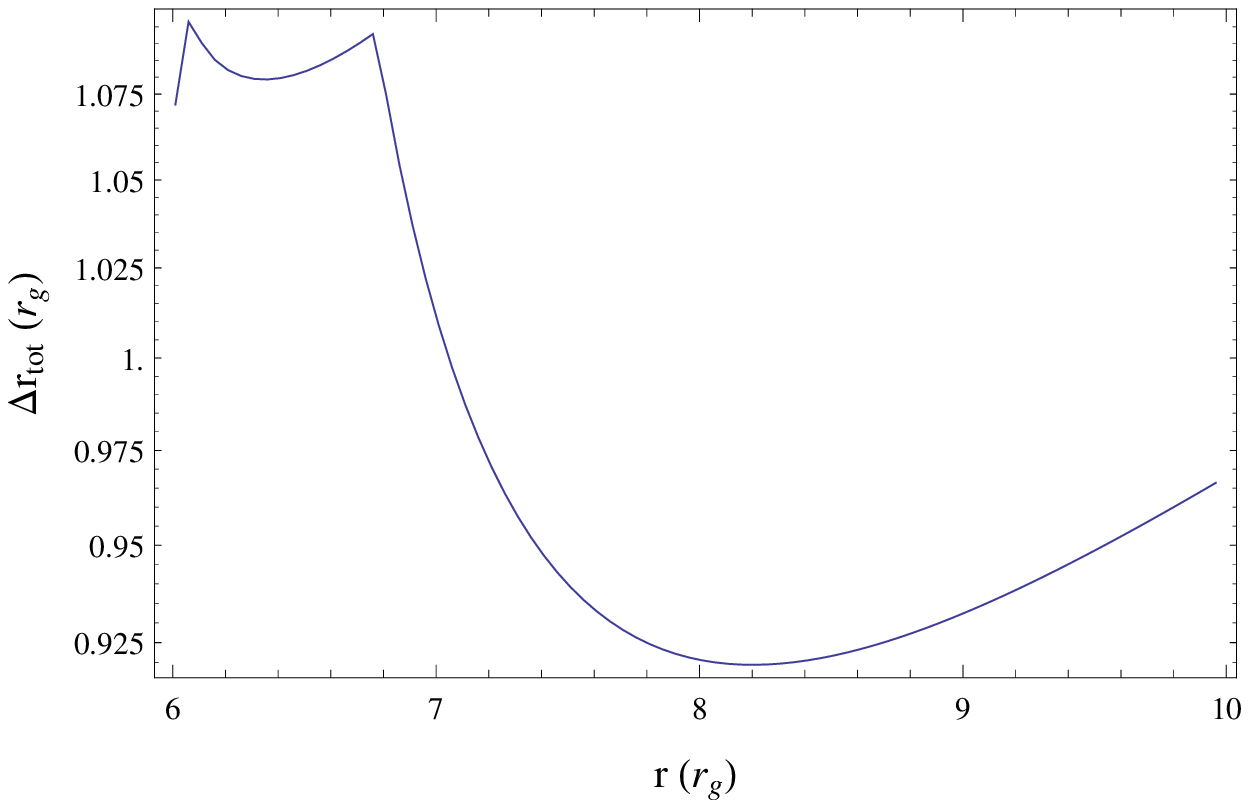}
\includegraphics[width=0.45\textwidth]{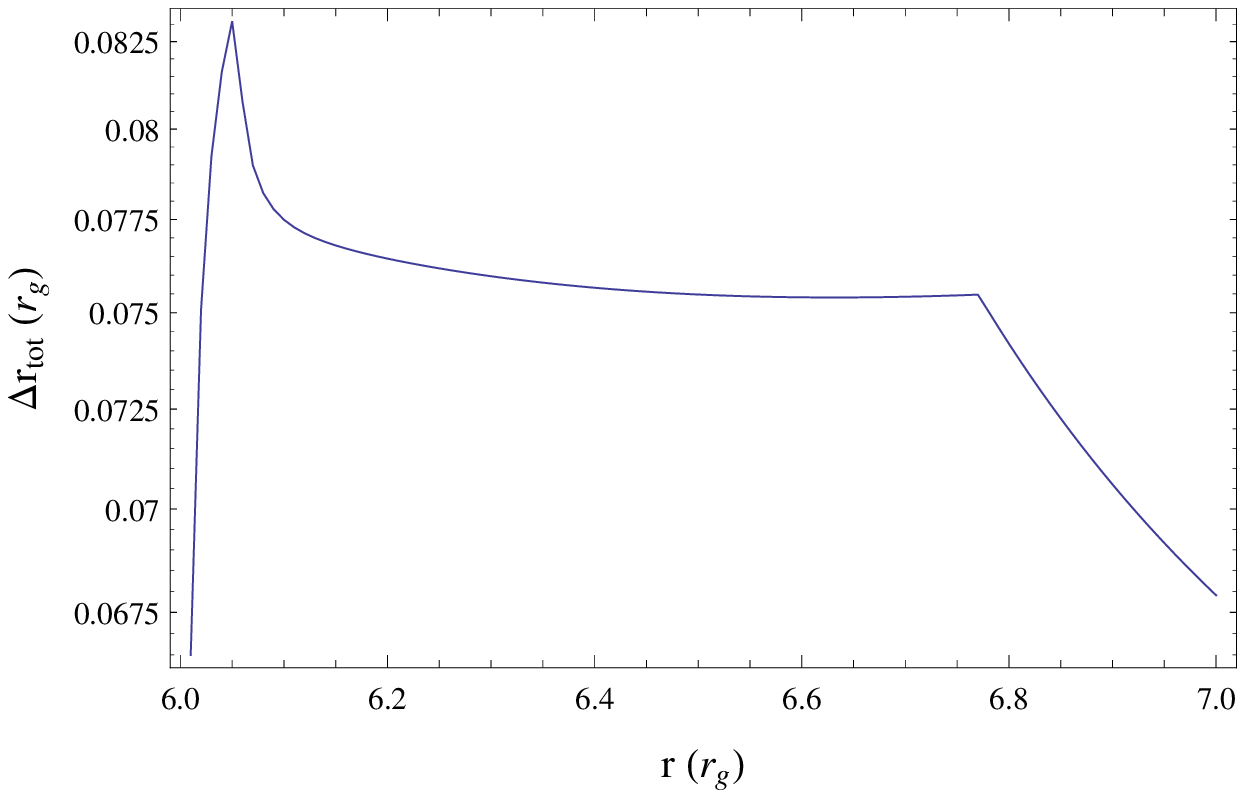}
\caption{Total radial drift $\Delta r_{tot}$ as a function of the orbital radius (units $r_{g}\sim 3$ km.) 
 \emph{Top}: $\Delta r_{tot}$ for $N_{cycle}=100$. \emph{Bottom}: $\Delta r_{tot}$ for $N_{cycle}=5$}\label{fig5}
\end{figure}
where $d\tilde{E}/dp$ is the first derivative of the orbital energy (\ref{eq8}) with respect to the semilatus 
 rectum $p$ of the orbit (since in this case $e=0$, $p=r/r_{g}$, and $r$ is radius of the circular orbit), 
 $E_{T}/\mu c^{2}$ is the tidal energy\footnote{Note that the order of magnitude obtained 
 $E_{T}\sim 0.01\%-0.1\%\ \mu c^{2}$ (the spheroidal-spherical clump; see also G17) agrees 
 with the tidal energy calculated with different formalisms \cite{2005ApJ...625..278G}.}
 $E_{T}=2\pi \sigma_{T}'V$ loaded on the spheroid 
 over one Keplerian turn, in units of its rest-mass energy. 
 The factor $(\tau/t_{k})^{-1}$ takes into account that the tidal wave propagates through the spheroid over 
 a finite time $\tau>t_{k}$, thus delaying the removal of orbital energy at each Keplerian turn.
 Fig.~\ref{fig4} shows $\Delta r$ (in units of $r_{g}\sim 3$ km) for three representative turns as a function 
 of the orbital radius $r$. The blue line is for Keplerian turn 1, 
 the orange for turn 50 and the green line for turn 100. Since in the simulation 
 we took into account that at each turn tides further squeeze the spheroid, for subsequent  
 turns the tidal energy is lower than previous turns. Thus, $\Delta r $ is smaller.
 This is seen in the figure because most of the 
 radial drift is just within turn 1. Turns 50 and 100 show a much lower radial drift. 
 The total radial drift $\Delta r_{tot}$ for a given number of turns, e.g. $N_{cycle}=100$, is shown in Fig.~\ref{fig5} (top)
 and it is the sum of the single radial drifts $\Delta r_{i}$ shown in Fig.~\ref{fig4}. We see that it increases 
 towards ISCO, reaching $\sim 1\ r_{g}$.  
The meaning of the figure is that close to or at ISCO ($r=6\ r_{g}$) an orbiting body will spiral on an inner unstable 
 orbit, soon falling onto the compact object. Thus, the signal emitted has a low temporal coherence, probably equal 
 to 0. Fig.~\ref{fig5} (bottom) shows that even a number of cycles $N_{cycle}=5$ gives a $\Delta r_{tot}\neq 0$
 for a body orbiting at ISCO, implying the body soon falling onto the compact object. We 
 can say that, within this framework, a body emitting radiation and orbiting close to or at ISCO will produce a Keplerian modulation 
 of temporal coherence essentially equal to 0.

To take into account the result in Fig.~\ref{fig5}, and link it to that shown in Fig.~\ref{fig3}, we construct 
 a table of values $(\nu_{k}, N_{cycle})$, i.e. Keplerian frequency $\nu_{k}$ versus number of cycles $N_{cycle}$. 
 Looking at Figs.~\ref{fig3},\ref{fig5} we see that for orbital radii $r< 7\ r_{g}$ the spiraling spheroid reaches 
 ISCO and, therefore, falls onto the compact object before be stretched by tides all over the orbit. Therefore, the 
 feature dictating the temporal coherence of the the Keplerian modulation is the number of Keplerian turns 
 to reach ISCO. Instead, for $r\ge 7\ r_{g}$ the spheroid is stretched by tides all over the orbit before reaching ISCO. Thus, 
 the temporal coherence of the Keplerian modulation produced is dictated by the tidal stretching timescale of the spheroid,   
 giving $N_{cycle}$ as in Fig.~\ref{fig3}. We then make an interpolation through the points of the table $(\nu_{k}, N_{cycle})$.    
\begin{figure}[!t!]
\includegraphics[width=0.45\textwidth]{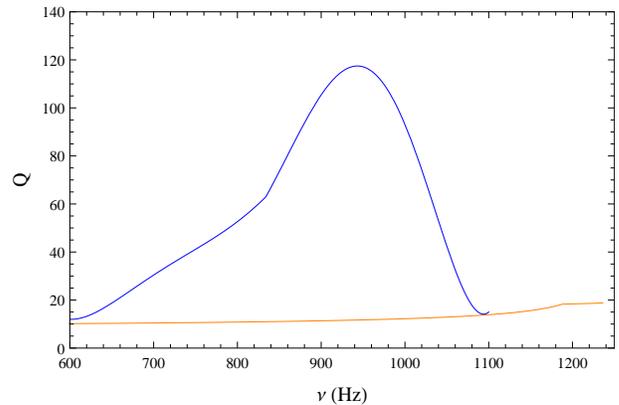}
\caption{Coherence $Q$ of both the Keplerian $\nu_{k}$ and $\nu_{k}+\nu_{r}$ modulations 
 produced by tidal disruption of magnetized clumps of plasma.
 \emph{Blue line}: coherence of the Keplerian modulation $\nu_{k}$. The increasing $Q$ is
 produced by the tidal stretching timescale of the spheroid deformed by tides. 
 The drop of $Q$ is dictated by the fall of the deformed clump of plasma onto the compact object.
 Such behavior is typical of the coherence of the lower HF QPO observed in several atoll NS LMXBs 
 (Fig.~2 in Ref.~\cite{2006MNRAS.370.1140B}). 
 \emph{Orange line}: coherence $Q$ of the beat $\nu_{k}+\nu_{r}$ produced by tidal circularization of relativistic 
 orbits (see G17). Such behavior is typical of the coherence of the upper HF QPO 
 observed in several atoll NS LMXBs (Fig.~2 in Ref.~\cite{2006MNRAS.370.1140B}).}\label{fig6}
\end{figure}

Figure \ref{fig6} shows the quality factor $Q$ of the Keplerian modulation produced by this physical
 mechanism\footnote{We report also the $Q$ of the beat $\nu_{k}+\nu_{r}$ obtained in G17 from tidal 
 circularization of the orbit. Here we corrected the trend for the finite propagation timescale 
 of the tidal wave through the clump, in the phase of tidal circularization of the orbit. It gives  
 a slight increase up to $Q\sim 20$ at high frequencies. Note that, as seen 
 in Fig.~\ref{fig2}, for $r\ge 10\ r_{g}$ ($\nu_{k}+\nu_{r}\leq 700$ Hz in Fig.~\ref{fig6})
 the clump is stretched by tides all 
 over the orbit before complete tidal circularization of the orbit. Therefore, the actual trend of $Q$ 
 for $\nu\le 700$ Hz should be lower than the orange line in Fig.~\ref{fig6}, as seen in the data, 
 where the $Q$ of the upper HF QPO steadily increases from $Q\sim 5$ to $Q\sim 20$
 over the range of frequency $600-1200$ Hz.}. 
 The quality factor $Q$ of twin peak HF QPOs is defined as $Q=\nu/\Delta\nu$ with $\nu$ central frequency
 of the peak in the power spectrum and $\Delta\nu$ its width at half maximum (e.g. Ref.~\cite{2006MNRAS.370.1140B}).
 For amplitude modulation, in Fourier temporal analysis $\Delta\nu$ is the inverse
 of the timescale the oscillation lasts, i.e. the coherence timescale, 
 $\Delta\nu\sim t'^{-1}$. Since $t'=t_{k}N_{cycle}$, with $t_{k}$ Keplerian period, 
 $Q=\nu_{k}/\Delta\nu=N_{cycle}$ we have derived here. From Fig.~\ref{fig6} we see that the coherence of 
 the Keplerian modulation $\nu_{k}$ is characterized by an increase up to $\nu_{k}\sim 950$ Hz
 (\footnote{Note that we have calculated the mean Keplerian frequency within the radial drift interval 
 $\Delta r_{tot}$ over which the clump spirals. Moreover, a boundary layer close to or at the NS surface may affect 
 the Keplerian frequency of the orbiting matter in the innerpart of the accretion disk, such that the real orbital 
 motion is not purely Keplerian. Also, we used the Schwarzschild metric to be consistent with the results from tidal
 circularization (G17 and Fig.~\ref{fig2} here), where we needed parametrizations of $\tilde{E}$ and $\tilde{L}$ 
 in the case of orbits with eccentricity $e\neq 0$ \cite{1994PhRvD..50.3816C}. 
 Thus, the orbital frequencies in Fig.~\ref{fig6} are an approximation.} $\sim 6.5\ r_{g}$) followed by an abrupt drop 
 towards ISCO\footnote{At ISCO $Q$ is not zero because we have added to the blue line 
 the coherence of $\nu_{k}$ produced in the phase of tidal circularization of the orbit
 ($Q\sim 15$, see G17).} ($\nu_{k}\sim 1100$ Hz, $r=6\ r_{g}$). The increasing $Q$ 
 is a consequence of the weakening of tides closer to ISCO. Also, it is because 
 the clump was less deformed during tidal circularization of the orbit, so needs to make more turns 
 to be stretched all over the orbit. Thus, the 
 tidal stretching timescale increases, giving a higher $N_{cycle}$.
 Weakening of tides towards ISCO (see also Fig.~2 in G17) is an effect of
 the flattening of the gravitational potential minimum, 
 a feature of a curved space-time and caused by the term $\propto 1/r^{3}$ in (\ref{eq2}) \cite{misner}. 
 The drop of $Q$ is produced by the fact that the spheroid reaches ISCO before being stretched by tides 
 all over the orbit and then falls onto the compact object on unstable orbits.\\
A feature that might affect the drop in Fig.~\ref{fig6} is the contribution 
 by the drifting frequency $\Delta\nu_{drift}$ during 
 spiraling. An attempt to estimate it is presented 
 in Ref.~\cite{2006MNRAS.370.1140B}, where $\Delta\nu_{drift}$ would sum quadratically to  
 $\Delta\nu$ due to the oscillation finite lifetime we have estimated here, i.e.  
 the tidal stretching timescale of the clump all over the orbit. 
 In Ref.~\cite{2006MNRAS.370.1140B} the contribution of $\Delta\nu_{drift}$ is  
 relevant close to ISCO, likely giving a sharper drop of $Q$ than that seen in Fig.~\ref{fig6}.\\    
The strength of the initial magnetic field $B$ permeating the clump might affect the slope of the increasing $Q$. 
 As pointed out in Sec.~\ref{sec2}, 
 we set $B=\mu/r^{3}$ to take into account that plasma orbiting closer 
 to the NS might be permeated by a stronger $B$. On the other hand, as said in Sec.~\ref{sec3},
 once tidal circularization of the orbit takes
 place the tidal energy deposited on the clump increases its $B$, getting equal over the 
 range $r\sim 6-10\ r_{g}$. A different initial $B$ affects only the amount of 
 tidal stretching of the clumps during circularization of the orbit. 
 That is, clumps orbiting farther away are more deformed
 than if we set a constant initial $B$ as a function of $r$. Therefore, 
 clumps orbiting at $r\ge 10\ r_{g}$ would not find themselves stretched all over the orbit as in Fig.~\ref{fig2} if the 
 initial $B$ is constant with $r$. We, however, can conclude that the increasing $Q$ seen in Fig.~\ref{fig6} is 
 due also to the weakening of tides approaching ISCO, as well as to the less energy deposited on the clump 
 during circularization of the orbits closer to ISBO, deforming the clump to a lesser extent. This last feature is also 
 a consequence of the flattening of the gravitational potential minimum.  
   
The overall behavior in Fig.~\ref{fig6} is typical of the lower HF QPO coherence observed in several atoll NS LMXBs
 (Fig.~2 in Ref.~\cite{2006MNRAS.370.1140B} and Fig.~1 in Ref.~\cite{2006MNRAS.371.1925M}). 
 The authors in Ref.~\cite{2006MNRAS.370.1140B} have already proposed 
 that the drop of $Q$ seen in the data might be caused by ISCO. Here and in G17 we have performed
 a consistent modeling that, although an approximated one because of the complexity of the problem,
 is able to reproduce for the first time the $Q$ of both the twin peak HF QPOs. 
 We have proposed and highlighted a physical mechanism, i.e. 
 the removal of the clump orbital energy by strong tides, 
 circularizing the clump's orbit and stretching it over the orbit.
 Following these results (Fig.~\ref{fig6}), 
 we emphasize the proposal that the drop of $Q$ of the lower HF QPO seen in atoll NS LMXBs could be a 
 candidate to disclose the ISCO predicted by GR in the strong field regime ($r\sim r_{g}$). 
 We add that the increase of $Q$ could be as well a candidate for a signature of ISCO. 
 We also highlight that Fig.~\ref{fig6} strongly suggests  
 a lower HF QPO seen in NS LMXBs corresponding to $\nu_{k}$, 
 while the upper HF QPO corresponds to $\nu_{k}+\nu_{r}$, 
 in agreement with numerical simulations \cite{2009AIPC.1126..367G,2013MNRAS.430L...1G} 
 and previous conclusions (GC15, G17).
 	
\section{Conclusions}\label{sec5}

The accretion disk in LMXBs might be characterized by clumps of plasma propagating throughout it. 
 It is worth mentioning that the authors in Ref.~\cite{2013Sci...339.1048C} have discovered
 large structures in the accretion disk of a LMXB. Accretion rate fluctuations
 that propagate in the disk are modeled to interpret the 
 temporal variability seen in BH LMXBs \cite{2016AN....337..385I}. 
 The possibility to have a magnetically confined massive clump of plasma 
 in the inner part of the accretion disk was pointed out in Ref.~\cite{1989Ap&SS.158..205H}.
 In this manuscript we have pursued the works presented in GC15 and G17, where 
 we explored the idea of treating a clump of plasma, disrupted by strong tides, 
 as characterized by an internal pressure $\sigma$. 
 In G17 we highlighted that such pressure could be a magnetic one and 
 derived a magnetic field typical of that in atoll and Z NS LMXBs 
 ($\sim 10^{8}-10^{9}$ G \cite{1999A&AT...18..447P}).
 These works were motivated because of the results shown by numerical simulations 
 on tidal disruption of clumps of matter around a compact object \cite{2009A&A...496..307K},
 producing power spectra much like those observed \cite{2009AIPC.1126..367G}, with the characteristic
 twin peak HF QPOs \cite{2004astro.ph.10551V}.

Having in mind the complexity of the problem, here we have attempted to study how a magnetized clump of plasma
 would react to tidal deformation. We performed a consistent modeling with the results from G17, 
 where we showed that the upper HF QPO could originate from tidal circularization of the
 clump's relativistic orbit. Here we took into account internal magnetic pressure 
 of the clump and investigated the subsequent evolution phase, when the clump is deformed by 
 tides into a spheroidlike object and orbits on a circular orbit,  
 undergoing more tides.

The results presented here (Fig.~\ref{fig6}) suggest that the strong tidal
 force by a compact object could play a relevant role in the physical mechanism
 producing the twin peak HF QPOs seen in NS LMXBs. TDEs are known around supermassive BHs, where 
 stars are disrupted by tides and emit energy as flares 
 \cite{2014ApJ...783...23G,2015Natur.526..542M,2015JHEAp...7..148K,2017MNRAS.468..783L}.
 The coherence $Q$ in Fig.~\ref{fig6} of the Keplerian
 modulation $\nu_{k}$ is typical of the lower HF QPO coherence seen 
 in the data \cite{2006MNRAS.370.1140B,2006MNRAS.371.1925M}.
 Although in an approximated way, this modeling is able to reproduce, for the first time, 
 the twin peak HF QPO coherences seen in atoll NS LMXBs.
 The increasing $Q$ of $\nu_{k}$ in Fig.~\ref{fig6} is drawn by the tidal stretching timescale of 
 the clump of plasma all over the orbit. Afterwards, the clump 
 would no longer produce Keplerian modulations. The drop of $Q$ is drawn by the number of turns
 a clump makes before reaching ISCO. 
 Subsequently, the clump enters unstable orbits and falls onto the compact object.
 Both the increase and decrease of $Q$ are caused  by the ISCO 
 predicted by GR. In the increasing part, clumps orbiting at inner radii undergo 
 weakened tides because of the gravitational potential minimum flattening, a feature of GR 
 caused by the term $\propto 1/r^{3}$ in (\ref{eq2}) \cite{misner}. Thus, the tidal stretching timescale increases 
 towards inner orbital radii and so the number of turns in which the clump can produce 
 Keplerian modulations. The increasing $B$ towards the NS might affect the slope of $Q$.
 In the decreasing part of $Q$, the number of turns before falling 
 beyond ISCO decreases for clumps orbiting closer to ISCO.
 
We conclude that the $Q$ of the lower HF QPO seen in atoll NS LMXBs \cite{2006MNRAS.370.1140B,2006MNRAS.371.1925M} 
 might be a potential candidate to disclose magnetized clumps of plasma both
 strongly stretched by tides and falling onto a NS.
  
\begin{acknowledgments}
This work was supported by the program PNPD/CAPES-Brazil.
\end{acknowledgments}

\bibliography{biblio.bib}

\end{document}